\documentstyle[12pt]{article}
\def\beq{\begin{equation}}
\def\eeq{\end{equation}}
\def\beqa{\begin{eqnarray}}
\def\eeqa{\end{eqnarray}}

\begin{document}
\title{ \bf {Meson Cloud and $SU(3)$ Symmetry \\
Breaking in Parton Distributions}}
\author{F. Carvalho\thanks{e-mail: babi@if.usp.br}, 
F. O. Dur\~aes\thanks{e-mail: fduraes@if.usp.br},
F.S. Navarra\thanks{e-mail: navarra@if.usp.br}, \\ 
M. Nielsen\thanks{e-mail: mnielsen@if.usp.br} \ and \
F. M. Steffens\thanks{e-mail: fsteffen@if.usp.br}\\
{\it Instituto de F\'{\i}sica, Universidade de S\~{a}o Paulo}\\
{\it C.P. 66318,  05315-970 S\~{a}o Paulo, SP, Brazil}} 
\maketitle
\vspace{1cm}
\begin{abstract}
We apply the Meson Cloud Model to the calculation of nonsinglet parton 
distributions in the nucleon sea, including the octet and the decuplet 
cloud baryon contributions. We give special attention to the differences 
between nonstrange and strange sea quarks, 
trying to identify possible sources of $SU(3)$ flavor breaking. A 
analysis in terms of the $\kappa$ parameter is presented, and we find that the 
existing $SU(3)$ flavor asymmetry in the nucleon sea can be quantitatively 
explained by the meson cloud. We also consider the $\Sigma^+$ baryon, 
finding similar conclusions.
\\
PACS numbers 14.20.Dh~~12.40.-y~~14.65.-q
\\

\end{abstract}

\vspace{1cm}

\section{Introduction}

The presence of a flavor  asymmetry in the light antiquark sea of the proton is 
now clearly established \cite{hawker,peng}. It can be expressed either in terms 
of the difference, $\Delta (x) = \overline d(x) - \overline u(x)$, or in terms 
of the ratio, $R(x) = \overline d(x) / \overline u(x)$. The fact that this 
difference is larger than zero (or that the ratio is larger than one) is usually 
referred to as $SU(2)$ flavor symmetry breaking in the proton sea.

We will discuss in this paper the nonperturbative origin of the
breaking of flavor symmetry, both at the $SU(2)$ and at the $SU(3)$ level. 
To this end, we will study the suppresion factor of $\overline u$ antiquarks
in the $SU(2)$ case, defined as

\beq
\kappa_{(2)} = \frac{\int_0^1 dx  x \overline{d} (x,\mu^2)}
{\int_0^1 dx x \overline{u} (x,\mu^2)},
\label{kappa2}
\eeq
and the suppression factor of strangeness in the $SU(3)$ case:
\beq
\kappa_{(3)} = \frac{\int_0^1 dx [x s(x,\mu^2) + x \overline{s} (x,\mu^2)]}
{\int_0^1 dx [x \overline{u} (x,\mu^2) + x \overline{d} (x,\mu^2)]}.
\label{kappa3}
\eeq
We notice that in the limit of exact $SU(2)$ ($SU(3)$) flavor symmetry 
$\kappa_{(2)} = 1$ ($\kappa_{(3)} = 1$). 
The CCFR collaboration has measured \cite{ccfr} 
$\kappa_{(3)} \simeq 0.37 \pm 0.05 $ ($0.477 \pm 0.05$) in a LO (NLO) QCD
analysis. Uncertainties apart, it is clear 
that there is a substantial violation of the $SU(3)$ flavor symmetry. 
In the nonstrange light antiquark sector, the use of the standard parametrizations
leads to $\kappa_{(2)} \sim  1.3$ \cite{grv,mrs}, 
indicating also a strong violation of the $SU(2)$ flavor symmetry in 
the proton sea. At the same time, the $SU(2)$ charge symmetry is believed to hold 
within the baryon octet, i.e., $\overline d(x) - \overline u(x)$ in the proton is 
equal to $\overline u(x) - \overline d(x)$ in the neutron. An interesting question 
is how $SU(3)$ charge symmetry is broken within the baryon octet. If the symmetry
were exact, it would mean, for instance, that $s(x) - \overline s(x)$ in the proton 
should be equal to $d(x) - \overline d(x)$ in the $\Sigma^+$. 
However, as calculated by the authors of Ref. \cite{cao,alb,boros}, this is not
the case, and in this work we also investigate the origins of the breaking of 
this symmetry. 

In QCD, exact $SU(3)$ symmetry implies that the $u$, $d$ and $s$ quarks 
have the same mass. Since the strange quark mass, $m_s$, is significantly larger 
than the up and down quark masses, the symmetry is only approximate. At 
the hadronic level, exact $SU(3)$ symmetry  also implies that the masses of  
baryons or mesons belonging to  the same multiplets are all equal. Clearly this 
is not the case and the masses within the baryon multiplets differ among 
themselves by more than $30 \%$.  The mass discrepancy is even more 
pronounced in the meson octet. 

Another consequence of the $SU(3)$ symmetry at the hadronic level is that the 
coupling constant in a generic 
baryon-baryon-meson ($g \overline{B} \gamma_5 B M$) vertex should be the same 
for all $B$, $\overline{B}$ and $M$. Since these three states together must form 
a $SU(3)$ singlet state, and the mesons are usually in octet states, it follows 
that the product of the two baryon representations 
must also be in a  $SU(3)$ octet state. Out of the ($\overline{B} B$) product 
$8 \times 8$, we get two distinct octets and therefore two independent coupling 
constants. This is the origin of the two $SU(3)$ constants, $F$ and $D$. 
When we consider some particular baryon-baryon-meson vertices, additional 
(Clebsch-Gordan) factors appear, so that the final couplings are different from
each other. However, exact $SU(3)$ symmetry 
imposes a well defined connections between them. Finally, analysis of experimental 
data determine the relation between $F$ and $D$ in terms of the parameter 
\cite{rat} 

\beq
\alpha_D = \frac{D}{D+F} \simeq 0.64
\label{alfa}
\eeq

We can make use of QCD Sum Rules (QCDSR) to calculate the above mentioned coupling 
constants \cite{suhong,miru}. In 
this approach we are able to identify the $SU(3)$ breaking sources affecting the 
couplings, which are mainly the  quark and hadron mass differences. 
The different values of the condensates and other QCDSR parameters also play an 
important role.   

As for the origin os the asymmetry in the light antiquark distributions,
there is now strong indications that part of the nucleon sea comes from 
fluctuations of the original nucleon into baryon-meson states, i.e., from the 
meson cloud \cite{tony83,ku,mt,cdnn}.  
The Meson Cloud Model (MCM) is dominated by hadronic quantities like hadron masses 
and coupling constants. 
This bridge between the physics of parton distribution and the conventional hadron 
physics may also help us, by connecting one with the other, to understand both 
$SU(3)$ symmetry breaking at the hadron and parton levels.

\section{Parton Distributions in the MCM}

In what follows, we show the meson-baryon Fock decomposition of the proton and of 
the $\Sigma^{+}$. In the case of the proton, most of the material has been already 
presented elsewhere \cite{ku,mt,cdnn}. We include it here just for completeness.  
Parton distributions in the $\Sigma^{+}$ hyperon have been discussed in 
\cite{cao,alb,boros}, and we will also address them in this work. This will enable 
us to make a close comparison between the proton and hyperon parton distributions.

\subsection{The proton}

As usual, we  decompose the proton in the following possible Fock states:
\beqa
|p> &=&  Z \, [ \,\, |p_0> +\,|p_0 \pi^0> +\, |n \pi^+> +\, |\Delta^0 \pi^+> +\, 
|\Delta^+ \pi^0> 
+\, |\Delta^{++} \pi^->  \nonumber\\  
& & +\, |\Lambda K^+> +\,  |\Sigma^0 K^+> +\,  |\Sigma^{0 *} K^+> +\, 
 |\Sigma^+ K^0> +\,  |\Sigma^{+ *} K^0> \,\, ] 
\label{fock}
\eeqa
where $|p_0>$ is the bare proton. We consider only light mesons. 
The relative normalization of these states 
is, in principle, fixed once the cloud parameters are given . The normalization
constant $Z$ measures the probability to find the proton in its bare state.

In the $ | MB > $ state,  the meson and the baryon 
have  fractional momentum $y_M$ and $y_B$, with distributions  
$f_{M/MB}(y_M)$ and $f_{B/MB}(y_B)$, respectively. Of course $y_M + y_B = 1$ and 
these distributions are related by:

\beq
f_{M/MB}(z) =  f_{B/MB}(1-z) 
\label{fmb}
\eeq

The splitting function $f_{M/MB} (y)$ represents the probability density to find 
a meson with momentum fraction $y$ of the nucleon and is usually given by 

\begin{equation}
f_{M/MB} (y) = \frac{g^2_{M B p}}{16 \pi^2} \, y \, 
\int_{-\infty}^{t_{max}}
dt \, \frac{[-t+(M_B-M_{p})^2]}{[t-m_{M}^2]^2}\,
F_{M B p}^2 (t)\; ,
\label{fpin}
\end{equation}
for baryons ($B$) belonging to the octet, and  

\begin{equation}
f_{M/MB} (y) = \frac{g^2_{M B p}}{16 \pi^2} \, y \, 
\int_{-\infty}^{t_{max}}
dt \, \frac{[(M_B + M_{p})^2 - t]^2 [(M_{p} - M_B)^2 - t]}
{ 12 M_B^2 M_{p}^2 [t-m_{M}^2]^2}\,
F_{M B p}^2 (t)\; 
\label{fpidel}
\end{equation}
for baryons belonging to the decuplet.
In the calculations we need the baryon-meson-baryon form factors 
appearing in the splitting functions. Following a phenomenological
approach, we use the dipole form:

\begin{equation}
F_{M B p} (t) = \left(  \frac{ \Lambda^2_{M B p} - m_{M}^2} 
                       {\Lambda^2_{M B p} - t} \right)^2
\label{eq:form}
\end{equation}
where $\Lambda_{M B p}$ is the  form factor cut-off parameter. 
In the above equations $t$ and $m_{M}$ are the four momentum square and the mass of
the meson in the cloud state, 
$t_{max}$ is the maximum $t$ given by:

\begin{equation}
t_{max} = M^2_p y- \frac{M^2_{B} y}{1-y}  \,\,\,\, , 
\label{tmax1}
\end{equation}
where $M_B$ ($M_{p}$) is the mass of the baryon (proton). Since the function 
$f_{M/MB} (y)$ has the interpretation of a flux of mesons inside 
the proton, the corresponding integral

\beq
n_{M/MB} = \sum_{MB} \,\,  \int_{0}^{1} \,\, d y \,\,f_{M/MB} (y),  
\label{nmeson}
\eeq
can be interpreted as the number of mesons in the proton,  or the number 
of mesons in the air. In many works, the magnitude of the multiplicities $n_{M/MB}$ 
has been considered as a
measure of the validity of MCM in the standard formulation with $MB$ states. If 
these multiplicities turn out to be large ( $\simeq 1$) then there is no 
justification for employing a one-meson truncation of the Fock expansion, as the 
expansion ceases to converge. This may happen for large cut-off values. 

Once the splitting functions (\ref{fpin}) and (\ref{fpidel}) are known we can 
calculate  the
antiquark distribution in the proton coming from the meson cloud through the 
convolution: 

\beq
\overline q_{f} (x) = \sum_{MB} \,\, \int_{x}^{1} \frac{d y}{y} f_{M/MB} (y)\,\, 
\overline q^{M}_{f} (\frac{x}{y})
\label{quark}
\eeq
where $\overline q^{M}_{f} (z)$ is the valence antiquark distribution of flavor $f$ 
in the meson. An analogous expression holds for the quark distributions. With the 
above formula we can compute the $\overline d$ and $\overline u$ distributions,
their  difference, $\overline d(x) - \overline u(x) $, and hence the Gottfried 
integral:

\beq
S_G = \frac{1}{3} - \frac{2}{3} \int_{0}^{1}  [ \overline d (x) - \overline u(x) ] dx 
\label{gottfried}
\eeq

Since we are interested in determining the sources of  $SU(3)$ symmetry  breaking, 
we also study the parton distributions in the  
case where the $SU(3)$ symmetry is exact. 
In our case, this is the limit in which we take all the meson and  baryon masses 
to be the same in the $SU(3)$ multiplets. All other ingredients are, from the 
start, compatible with $SU(3)$ symmetry, i.e.,  all coupling constants follow 
$SU(3)$ relations \cite{swart}, and the  cut-off parameters are the same for a 
given multiplet. Of course, the nonstrange subset of these couplings respects 
the $SU(2)$ (isospin) symmetry. 

The masses are $m_{p}=m_{n}=938\; MeV$, $m_{\pi}=138\; MeV$, $m_{K} = 480\; MeV$, 
$m_{\Delta}= 1232\; MeV$, $m_{\Lambda}=1116\; MeV$, and $m_{\Sigma}=1189\; MeV$. 
The octet  coupling  constants are given by the expressions in Table I \cite{swart}, 
where $ g_{p \pi^0 p} = -13.45$ \cite{ingelman,timm}
and $\alpha_D$ was given in eq. (\ref{alfa}).  For the 
decuplet coupling constants, in Table II, where  $g_{p \Delta^0 \pi^+}= 
\frac{28.6}{\sqrt{6}}$ \cite{werner,muko}, we also use the standard $SU(3)$ relations 
between the couplings \cite{swart}.

\begin{center}
\begin{tabular}{|c|c|}  \hline
$g_{p K^+ \Lambda}$  & $ - \frac{1}{\sqrt{3}}\,(3 - 2 \alpha_D) g_{p \pi^0 p} $ \\
\hline
$g_{p K^+ \Sigma^0} $  & $  (2 \alpha_D - 1) g_{p \pi^0 p}   $ \\
\hline
$g_{p K^0 \Sigma^+} $  & $ \sqrt{2} (2 \alpha_D - 1) g_{p \pi^0 p}   $ \\
\hline
$g_{p \pi^+ n} $  & $ \sqrt{2} g_{p \pi^0 p}   $ \\
\hline
\hline
\end{tabular}
\end{center}
\begin{center}
{\bf Table I:}  Octet coupling constants.
\end{center}

\begin{center}
\begin{tabular}{|c|c|}  \hline
$g_{p \Sigma^{*0} K^+}$  & $  \frac{\sqrt{2}}{2} \, g_{p \Delta^0 \pi^+} $ \\
\hline
$g_{p \Sigma^{*+} K^0}$  & $ g_{p \Delta^0 \pi^+} $ \\
\hline
$g_{p \Delta^+ \pi^0}$ & $ \sqrt{2}  g_{p \Delta^0 \pi^+}$ \\
\hline
$g_{p \Delta^{++} \pi^-}$ & $ \sqrt{3}  g_{p \Delta^0 \pi^+}$ \\
\hline
\hline
\end{tabular}
\end{center}
\begin{center}
{\bf Table II:}  Decuplet coupling constants.
\end{center}

Many of the works \cite{ku,mt,cdnn} done so far on this subject indicate that 
the cut-off parameters must be soft ($\Lambda \simeq 1$ GeV in dipole form). Indeed, in
our attempt to have a simultaneous description of both the difference 
$\Delta(x)$ and the ratio $R(x)$, we find that  
 
\beqa 
\Lambda_{oct} = 1.11 \,\,  GeV  \,\,\,\,;\,\,\,\, \Lambda_{dec} = 1.07 \,\,  GeV \,\, ,
\label{lambda}
\eeqa
where $\Lambda_{oct}$ and 
$\Lambda_{dec}$ are the cut-off parameters for all the octet and decuplet vertices 
respectively. We notice that the ratio and the difference are related by \cite{hawker}:

\beq
R(x) = \frac{1 + \Delta (x)/\Sigma (x)}{1 - \Delta (x)/\Sigma(x)},
\label{Ratiox}
\eeq
where $\Sigma (x) = \overline{d}(x) + \overline{u}(x)$ is the total distribution, 
which may be taken from any of the available parametrizations of the parton 
distributions. We see from the dashed lines of Figs. 1a and 1b, that {\it it is 
not possible} to have a simultaneous description of the ratio and of the 
difference using the MCM only. Additional, nonperturbative physics is necessary, 
and we will discuss this point below. 
Before that, we point out that it is crucial for our discussion of $SU(3)$ symmetry 
breaking that the cut-off parameters be the same  
for all the members of the multiplets, including the cut-offs involved
in the production of strangeness. Their exact values could be different, provided 
that all the constraints imposed by convergence of the Fock expansion, by data on 
inclusive meson production or any other experimental information, be satisfied.
In any case, we stress that the whole set of cut-offs were fixed in the reproduction
of the E866 data, and the results presented here, which includes the strange sector,
are predictions of the model.

As we noticed before, there should be some extra, nonperturbative, physics in order
to describe the full E866 data points. The natural step to take is to consider 
effects from the Fermi statistics of the quarks, as suggested long ago 
by Field and Feynman \cite{field}, and implemented recently in a quantitative
way \cite{mt}. The idea is quite straight: as the proton is, primaraly, a $uud$ 
state, 
it should be easier to insert a $d\overline d$ pair than a $u\overline u$ pair
in the proton sea. This follows from the fact that there are more empty states
for the insertion of a $d$ quark than for the insertion of a $u$ 
quark\footnote{As shown in \cite{dg,me}, possible antisimmetrization effects 
between the sea and the valence quarks can spoil this naive counting.}.
Following Ref. \cite{mt}, we parametrize this Pauli Blocking (PB) contribution by:

\beq
(\overline{d} -\overline{u})^{\mbox{PB}} (x)= \Delta^{\mbox{PB}} (n+1) (1 - x)^n.
\label{pauli}
\eeq
As part of the nonperturbative
sea, the PB contribution is added to the $\overline d(x) -\overline u(x)$ 
difference coming from the meson cloud, computed from Eq. (\ref{quark}). 

The dashed lines in Figs. 1a and 1b show the combined effect of meson cloud
and PB effects for the $\overline d(x) - \overline u(x)$ and for 
$\overline d(x)/\overline u(x)$, as a function of $x$, using $\Delta^{\mbox{PB}} 
= 0.017$ and $n=10$. The data points are from the
E866  collaboration \cite{hawker,peng}, where the CTEQ parametrization for 
$\overline u(x) + \overline d(x)$ was used in Eq. (\ref{Ratiox}) to relate
$\Delta (x)$ to $R(x)$. Our results confirm analogous calculations performed 
previously by Melnitchouk, Speth and Thomas \cite{mt}, although we see that 
the size of our PB is significantly smaller.  

When working with the sea parton distributions it should be emphasized that in 
differences such as $\overline d(x) - \overline u(x)$  or 
$\overline s(x) - \overline u(x)$ 
the perturbative contributions should cancel if the production of sea partons 
from hard gluons is to be insensitive to small masses, including the strange 
quark mass.  
Such a property was already used in the writing of Eq. (\ref{Ratiox}).
Therefore, any deviation of $\kappa_{(2)}$ and $\kappa_{(3)}$ from 
1 (or $x(\overline d(x) - \overline u(x))$, etc, from zero), must have a 
nonperturbative origin. As the meson cloud is the main nonperturbative 
contribution, it should be quite reliable when calculating
the differences of sea distributions. Figure 1 supports this view.  
The ratios, on the other hand, also include the (dominant) perturbative 
contribution.  
Thus, in order to calculate $\kappa_{(2)}$ including this contribution, we use
the fact that $\int_0^1 x (\overline d(x) - \overline u(x))dx \neq 0$ from 
nonperturbative effects only, and rewrite Eq. (\ref{kappa2}) as:

\beq
\kappa_{(2)} = 1 + \frac{\int_0^1 dx x [\overline d(x) - \overline u(x)]^{NP}}  
{\int_0^1 dx x \overline u(x)},
\label{newkappa2}
\eeq
where in the denominator we have used the CTEQ4 parametrization \cite{cteq}
for the integral of the $\overline u$ antiquark distribution. 
The obtained value is
\beq
\kappa_{(2)} = 1.22,
\eeq
compatible with the values quoted in the introduction. 

For the Gottfried sum rule (\ref{gottfried}), we obtain $S_G = 0.255$, which is to 
be compared with the experimental value $0.235 \pm 0.026$, obtained by the E866 
collaboration \cite{hawker,peng}. The calculation of the multiplicities through 
Eq. (\ref{nmeson}) give $n_{\pi N} \simeq 0.30$ and $n_{\pi \Delta} \simeq 0.27$.

Before moving to the strange sector, it is worth noticing that this value of 
$\kappa_{(2)}$ indicates a violation of $SU(2)$ flavor inside the 
proton which is not in conflict with the $SU(2)$ charge symmetry between the 
proton and 
the neutron. The $SU(2)$ charge symmetry still holds in the 
MCM. In order to check this, it is enough to 
write the dominant terms of the Fock expansion for the neutron cloud
\beqa
|n> &=&  Z \, [ \,\, |n_0> +\, |n_0 \pi^0>  + \, |p \pi^->  +\, |\Delta^0 \pi^0> +\, 
|\Delta^+ \pi^-> 
+\, |\Delta^{-} \pi^+>  \,\, ], 
\label{fockn}
\eeqa
and realize that, since our coupling constants respect $SU(2)$, it follows that  
$g_{n p \pi^-} = g_{p n \pi^+}$, 
$g_{n \Delta^-  \pi^+} = -  g_{p \Delta^{++}  \pi^-}$, and 
$g_{n \Delta^+ \pi^-} = - g_{p \Delta^0 \pi^+}$. When we substitute these relations
in Eq. (\ref{quark}), and use $m_p = m_n$, we arrive at the conclusion that 
$\overline d(x) -\overline u(x)$ in the proton is exactly the same as 
$\overline u(x) -\overline d(x)$ in the neutron.

In Figs. 2a, 2b and 2c we show, respectively, $x s(x)$ (compared to $x \overline{s}$),
$x (s(x) - \overline s(x))$ (decomposed in its octet and decuplet contributions),
and  $s(x) - \overline s(x)$ as functions of $x$. We made the assumption 
that the valence $s(x)$ distribution in the 
hyperons is the same as the valence $d(x)$ distribution in the proton. 
The $s(x)$ quark distribution is harder than the $\overline s(x)$ distribution 
because it is inside a (harder) strange baryon in the cloud,
a conclusion corroborated by other authors \cite{boma,mane}. 
In Fig. 2b  we can appreciate how significant the decuplet contribution is for the
$x(s(x) - \overline s(x))$ asymmetry, which is plotted in Fig. 2c against the 
experimental result (shaded area) \cite{ccfr}.

In Fig. 3 we show  $\overline s(x) - \overline u(x)$ in the proton. 
The points represent the parametrizations CTEQ4L (full circles), GRV98LO 
(open circles) and MRS99-1(squares) and the solid line shows the MCM result. 
The dashed and dotted lines show the octet and decuplet contributions to 
the meson cloud, respectively. From this figure we can see that the decuplet 
contribution is sizeable, and therefore it should be included in 
any study of nonsinglet quantities involving the strange sea quark distributions. 

When extending our analysis from 2 to 3 flavors, we can define a quantity
analogous to the $\overline d(x) - \overline u(x)$ difference, i.e., 
a quantity which measures how blocked is the production of strange
quarks compared to the nonstrange quarks: $\overline d(x) + \overline u(x) - 
s(x) - \overline s (x)$. Notice that, from the point of view of perturbative QCD,
this quantity should be zero (besides, perhaps, some small mass effects, which
should not be relevant in the intermediate or small $x$ region). Hence, if
our current view of the nonperturbative proton sea, as generated from mesons
and from Pauli blocking, is correct this difference should also be well described
by the MCM. We show our results in Fig. 4, and we see that our MCM curve (solid 
line) is
compatible (although on the edge) with the values extracted from the different 
parametrizations for the parton distribution functions. 
We then conclude that the proportion of strange to non-strange quarks as 
calculated in the MCM is compatible with what the standard parametrizations
for parton distributions tell us. For illustration, we also show in the 
dashed lines what would be $\overline d(x) + \overline u(x) - s(x) - \overline s (x)$ 
in the $SU(3)$ symmetry limit, which will be defined in Eq. (\ref{su3lim}).

The combination of 
parton distributions shown in Fig. 4 is useful for the computation of the factor 
$\kappa_{(3)}$. Indeed, the 
numerator and denominator of Eq. (\ref{kappa3}) can be rewritten, as before, as sums of 
a perturbative $(P)$ plus a non-perturbative $(NP)$ contributions: 

\beq
\int_0^1 dx x [ s + \overline{s}](x) = \int_0^1 dx x [ (s + \overline{s})^{P} + 
(s + \overline{s})^{NP}](x),
\label{ssbar}
\eeq
\beq
\int_0^1 dx x [\overline{u} + \overline{d}](x) = \int_0^1 dx x 
[ (\overline{u} + \overline{d})^{P} + (\overline{u} + \overline{d})^{NP}](x).
\label{ubdb}
\eeq
Subtracting (\ref{ubdb}) from the (\ref{ssbar}), dividing both sides by (\ref{ubdb}), 
and assuming that all the perturbative contributions cancel in the numerator we 
rewrite $\kappa_{(3)}$ as:

\beq
\kappa_{(3)} = 1 + \frac{\int_0^1 dx x [ s + \overline{s}]^{NP}(x) -  
\int_0^1 dx x [\overline{u} + \overline{d}]^{NP}(x)}  
{\int_0^1 dx x [\overline{u} + \overline{d}](x) },
\label{newkappa}
\eeq
where, as in Eq. (\ref{newkappa2}), we have used the CTEQ4 parametrizations in the 
denominator. In the above expressions the non-perturbative quantities are 
calculated with the MCM. Using the parameters described before we find

\beq
\kappa_{(3)} = 0.55,
\eeq
in reasonable agreement with the value quoted by the CCFR collaboration 
\cite{ccfr}. 

The cloud parameters used so far give an overall good agreement with the 
available experimental information. However, they are not the result of a best 
fit, and a different set of parameters could yield good results as well. In 
particular, we would like to mention that our value for $g_{p \Delta^0 \pi^+}$  
is somewhat large (although still compatible with
data) and, as it was argued in \cite{mt}, a value about $30\%$ smaller might be 
more appropriate. We repeated our calculation using $g_{p \Delta^0 \pi^+}= 
\frac{22.0}{\sqrt{6}}$. The cut-off parameters had to be changed to 
$\Lambda_{oct} = 1.11$ GeV and $ \Lambda_{dec} = 1.15$ GeV, and the new 
multiplicities were 
calculated to be $ n_{\pi N} \simeq 0.30$ and $ n_{\pi \Delta} 
\simeq 0.19$. On the other hand,  $\kappa_{(3)} = 0.66$ with this new set 
of parameters, implying in less agreement between the model and the
experimental data.

We now take the $SU(3)$ symmetry limit, which means in our case to  
make the masses equal within the multiplets\footnote{Other choices for the
values of the masses in the symmetry limit would, of course, result in 
a different value for $\kappa_{(3)}$, for instance. The important point 
is that equal masses within the multiplets, indicate a tendency to 
recover the $SU(3)$ symmetry.}, i.e., 

$$
m^{octet}_{meson} = (m_{\pi} + m_K)/2, 
$$
\beq
m^{octet}_{baryon} = (m_p + m_n + m_{\Sigma} + m_{\Lambda})/4,
\label{su3lim} 
\eeq
$$
m^{decuplet}_{baryon} = (m_{\Sigma^{*}} + m_{\Delta})/2.
$$

As $\kappa_{(3)}$ in Eq. (\ref{newkappa}) measures the amount of symmetry 
breaking between the strange and nonstrange quarks, it is 
remarkable that whithin our $SU(3)$ symmetry limit, we have $\kappa_{(3)} = 0.96$,
which is in good agreement with $\kappa_{(3)} = 1$. We see, 
therefore, that in making the cloud $SU(3)$ symmetric, we recover the $SU(3)$ 
flavor symmetry in the parton  distributions.

It is of capital importance to compare the $SU(3)$ symmetry limit, as defined
by Eq. (\ref{su3lim}), with a similar limit in the $SU(2)$ case. Notice that
to calculate the $\overline d(x) - \overline u(x)$ difference, we are already 
using a limit similar to that of Eq. (\ref{su3lim}). That is, we have only
one mass in the meson octet, $m_{\pi}$, only one mass in the 
barion octet, $m_p = m_n$, and only one mass in the barion decuplet, 
as only the $\Delta$ is relevant in that case. The bulk of the 
$\overline d(x) - \overline u(x)$ difference comes, then,  from the mass
diffence {\it between} the octet and decuplet barion masses. We have checked
that when $m_p \sim m_\Delta$, $\kappa_{(2)} \sim 1$.
For the $\overline d(x) + \overline u(x) - s(x) - \overline s(x)$ difference,
however, the important contribution comes from the mass differences
{\it inside} the octet and decuplet states.


\subsection{The Sigma}

For the $\Sigma^{+}$ baryon we consider the following expansion:
\beqa
|\Sigma^{+}> &=&  Z \, [ \,\, |\Sigma^{+}_{0}> +\,|\Sigma^{+} \pi^0> +\, 
|\Sigma^{0} \pi^+> +\, |\Lambda^0 \pi^+> +\, |p \overline{K^0}> 
+\, |\Xi^{0} K^{+}>  \nonumber\\  
& & +\, |\Delta^{++} K^-> +\,  |\Delta^+ \overline{K^{0}}> +\,  
|\Sigma^{* +} \pi^{0}> 
+\,  |\Sigma^{* 0} \pi^{+}> +\,  |\Xi^{* 0} K^{+}> \,\, ]. 
\label{focksig}
\eeqa

We included the $|\Xi^{0} K^{+}>$ and  the  decuplet states  in 
the second line of Eq. (\ref{focksig}). These states were not  considered in 
\cite{cao}, and the authors of Ref. \cite{boros} considered only the
two lowest lying decuplet states ($\Sigma^{* +} \pi^{0}, \Sigma^{* 0} \pi^{+}$).
It will be seen here that the decuplet states play an important role in the $x$ 
dependence of the parton distributions, in spite of their large masses. 

The parton distributions in the $\Sigma^{+}$ sea can be straightforwardly computed 
through Eqs. (\ref{fmb})-(\ref{quark}), where the relevant replacements of masses 
and couplings have to be made. 
Following the steps of subsection 2.1, we take the couplings according to the 
$SU(3)$ relations \cite{swart}. Hence, for the octet coupling constants we have:
\begin{center}
\begin{tabular}{|c|c|}  \hline
$g_{\pi^+ \Lambda \Sigma^+}$  & $\frac{2}{\sqrt{3}}\,2 \alpha_D \, g_{p \pi^0 p} $ \\
\hline
$g_{K^+ \Sigma^+ \Xi^0}$  & $ - g_{p \pi^0 p} $ \\
\hline
$g_{\Sigma^+ \Sigma^+ \pi^0}$  & $ 2 ( 1- \alpha_D) g_{p \pi^0 p} $ \\
\hline
$g_{p \overline{K^0} \Sigma^+} $  & $ \sqrt{2} (2 \alpha_D - 1) g_{p \pi^0 p}   $ \\
\hline
$g_{\Sigma^+ \pi^+ \Sigma^0} $  & $  2 ( 1- \alpha_D) g_{p \pi^0 p}    $ \\
\hline
\hline
\end{tabular}
\end{center}
\begin{center}
{\bf Table III:}  $\Sigma^{+}$ Octet coupling constants.
\end{center}

while for the decuplet couplings we have:
\begin{center}
\begin{tabular}{|c|c|}  \hline
$g_{\Sigma^+ \Delta^+ \overline{K^0}}$  & $ g_{p \Delta^0 \pi^+} $ \\
\hline
$g_{\Sigma^+ \Delta^{++}  K^-} $  & $  \sqrt{3} \, g_{p \Delta^0 \pi^+}  $ \\
\hline
$g_{\Sigma^+ \Xi^{*0} K^+}$  & $  g_{p \Delta^0 \pi^+} $ \\
\hline
$g_{\Sigma^+ \Sigma^{*0} \pi^+}$  & $ - \frac{1}{\sqrt{2}} \, g_{p \Delta^0 \pi^+} $ \\
\hline
$g_{\Sigma^+ \Sigma^{*+} \pi^0}$  & $ - \frac{1}{\sqrt{2}} \, g_{p \Delta^0 \pi^+} $ \\
\hline
\hline
\end{tabular}
\end{center}
\begin{center}
{\bf Table IV:}  $\Sigma^{+}$  Decuplet coupling constants.
\end{center}
For the cut-off parameters, we will use the same values as given by Eq. (\ref{lambda}).

In Fig. 5 we show the separate contributions from the octet and decuplet states
for $x (\overline{d}(x) -\overline{u}(x))$ (5a),  
$x (\overline{d}(x) -\overline{s}(x))$ (5b), 
$x (\overline{u}(x) -\overline{s}(x))$ (5c). 
The total distributions are shown in the Fig. 5d, and they should be 
compared with Fig. 4 of Ref. \cite{cao}. We agree qualitatively with them. 
Quantitative changes are noticeable, and they happen because of the 
inclusion of the decuplet states which play a significant role, as 
seen in Figs. 5b and 5c. The fact that 
$x (\overline d(x) -\overline u(x)) \,\, > \,\, x (\overline u(x) -\overline s(x))$, 
was interpreted in \cite{cao,boros} as a violation of $SU(3)$ charge symmetry, 
and this really seems to be the case. Even more indicative
of this breaking is the direct  comparison of $x (\overline d(x) -\overline u(x))$ 
in the proton (dotted line) with $x (\overline s(x) -\overline u(x))$ in the 
$\Sigma^+$ (dot-dashed line), shown in Fig. 6.
A huge discrepancy is seen between the two curves, a result in complete 
disagreement with naive expectations. As in the quark model the $\Sigma^+$ is a
proton with the $d$ quark replaced by a $s$ quark, naively one would think
that $x (\overline d(x) -\overline u(x))$ in the proton is equal to 
$x (\overline s(x) -\overline u(x))$ in the $\Sigma^+$.

As we saw in section 2.1, the PB effect is important in describing the 
$x$ dependence of the light quark sea asymmetry. From the point
of view of Fermi statistics, the same effect should be present in the $\Sigma^+$, 
with the $s$ quark here playing the role of the $d$ quark in the proton. Because 
of the mass of the $s$ quark, the $x$ dependence of the PB in the 
$\Sigma^+$ may not be exactly the same as in the proton. However, to 
exemplify the size of the corrections from PB, we also plot in Fig. 6
the distributions including the effect of the PB given by Eq. (\ref{pauli}).
The solid line is for $x (\overline d(x) -\overline u(x))$ in the proton, and 
the dashed line is for $x (\overline s(x) -\overline u(x))$ in the $\Sigma^+$.

It seems also appropriate to extend the comparisons to $d(x) - \overline d(x)$
in the $\Sigma^+$, and to $s(x) - \overline s(x)$ in the proton. We show the 
$d(x) - \overline d(x)$ in Fig. 7, where the decuplet and octet contributions
are shown separately. In Fig. 8 we show both differences and we see clearly 
the discrepancy between them, which is again an evidence of $SU(3)$ charge
symmetry breaking. 
It is remarkable, however, that besides the small mass of the 
$d$ quark, the $d(x) - \overline d(x)$ asymmetry in the $\Sigma^+$ is
much larger than the $s(x) - \overline s(x)$ asymmetry in the proton.

Finally, 
in order to compare the $SU(3)$ flavor breaking in sea parton distributions in the  
$\Sigma^{+}$ with the proton, we compute $\kappa_{(3)}$ defined 
in Eq. (\ref{newkappa}). The denominator in Eq. (\ref{newkappa}) is governed by the 
large perturbative contributions and is only slightly affected by the cloud 
component. It is therefore reasonable to assume that it is the same for the proton 
and for the $\Sigma^{+}$. In the numerator we have approximated  
$\int dx x [ s + \overline{s}]^{NP}$ by $\int dx x [ 2 \, \overline{s}]^{NP}$ in 
order to avoid uncertainties associated with $s(x)$ in the hyperon. 
The resulting value for $\kappa_3$ is then:

\beq
\kappa_3 \simeq 0.85
\eeq

This value of $\kappa_3$ indicates a violation of $SU(3)$ flavor inside the 
$\Sigma^+$ which is weaker than that inside the proton, whereas Figs. 4 - 7 
show a violation of the $SU(3)$ symmetry between the proton and the sigma. Both 
symmetries are restored in the $SU(3)$ symmetry limit of Eqs. (\ref{su3lim}), 
i.e.,  $\kappa_3 \rightarrow 1$ and 
the curves in the figures assume their expected behaviour, with 
${d}_{\Sigma^+} = {s}_p$ and
$ \overline{d}_{\Sigma^+} = \overline{s}_p$.
 
In the context of the meson cloud model this result is not surprising. The cloud 
expansion of the $\Sigma^{+}$ involves heavier states than those appearing in the
proton expansion. As a consequence, the whole $\Sigma^+$ cloud will be suppressed 
with respect to the proton cloud. Indeed, looking at the multiplicities 
we observe that the probabilities associated with the hyperon states are typically 
one order of magnitude smaller than those associated with the proton states.
Moreover, the strange states inside the proton are heavier and suppressed with respect to
non-strange states, and therefore we expect (and really observe) 
$ \overline{d} > \overline{u} > \overline{s} $. Neglecting Pauli blocking effects, 
(which would  slightly inhibit the $\overline{s}$ production in comparison with
the $\overline d$ production in the $\Sigma^+$) we 
would expect the same behaviour for the $\Sigma^+$ and this is exactly what we 
find. Quantitatively, the suppression of $ \overline{s}$ in $\Sigma^+$ (with 
respect to $ \overline{d}$ or $ \overline{u}$), happens  because all the states 
in the cloud contain strangeness and are nearly equally suppressed. In the proton 
the suppression of $ \overline{s}$ (always with respect to  
$ \overline{d}$ or $ \overline{u}$ ) is more pronounced because of the mass 
difference between strange and non-strange cloud states. 

\section{Conclusions}

In this work we have applied to meson cloud model to study the non-perturbative 
aspects of parton distributions, giving special emphasis to the strange sector. 
We have adjusted the cloud cut-off parameters to reproduce the E866 data on 
$\overline d(x) - \overline u(x)$ and $\overline d(x) / \overline u(x)$.  
In this procedure the choices were not completely 
free. Instead, the cut-off values had to be consistent with previous analises of 
other experimental information \cite{cdnn}. 
Having fixed the parameters we moved to the strange sector. In this sense, the
results for the strange-anti-strange asymmetry and for 
$\overline{u} + \overline{d}- s - \overline{s}$ can be considered as
predictions. They are consistent with data. Finally we have taken the $SU(3)$ 
limit in the  meson cloud and  found out that, in this limit, the parton 
distributions become $SU(3)$ flavor symmetric, i.e.,
$\kappa \rightarrow 1$. We have thus presented additional experimental confirmation 
of the MCM. Moreover we have concluded that the meson cloud is responsible for the 
$SU(3)$ flavor breaking in parton distributions.
                      
\vspace{1.0cm}

\underline{Acknowledgements}: This work has been supported by CNPq and  
FAPESP under contract number 98/2249-4. We are indebted to C. Shat, W. Melnitchouk
and A.W. Thomas for discussions.   

\vspace{1cm}
\newpage
\noindent
{\bf Figure Captions}\\
\begin{itemize}

\item[{\bf Fig. 1}] a) $\overline d(x) - \overline u(x)$ calculated with Eq. 
(\ref{quark}) compared with E866 data; b) same as a) for the ratio 
$\overline d(x)/\overline u(x)$. The
dashed lines represent our result without  Pauli blocking.

\item[{\bf Fig. 2}] a) $ x \overline s(x)$ (solid line) and $x s(x)$ (dashed line) 
in the proton computed with the  MCM (using Eq. (\ref{quark})); 
b) $x (s(x) - \overline s(x))$ in the proton in the MCM. The octet and decuplet 
contributions are represented by the dashed and the dotted lines, respectively; 
c) same as b) for the difference
$s(x) - \overline s(x)$. The shaded area is the uncertainty range ofthe 
experimental data \cite{ccfr}. 

\item[{\bf Fig. 3}] $ \overline s(x) - \overline u(x)$ in the proton extracted from
several parametrizations,  and the resulting curves from the  
MCM (solid line). The octet and decuplet contributions are 
the dashed and the dotted lines, respectively.

\item[{\bf Fig. 4}] a) $ \overline u(x) + \overline d(x) - s(x) - \overline s(x)$ 
in the proton extracted from several parametrizations, and the result from the  
MCM (solid line). The dashed line is the MCM in the $SU(3)$ limit.

\item[{\bf Fig. 5}] a) $x (\overline d(x)  - \overline u(x))$ in the $\Sigma^+$ 
calculated with the  MCM (solid line). The octet and decuplet contributions are 
the dashed and dotted lines, respectively; b) same as a) for 
$x (\overline d(x)  - \overline s(x))$; c) same as a) $x (\overline u(x)  - 
\overline s(x))$; d) All the curves together, where the decuplet and 
octet contributions were added.

\item[{\bf Fig. 6}] $x (\overline d(x)  - \overline u(x))$ in the proton, with 
(solid line) and without (dotted line) Pauli blocking. 
 $(\overline s(x)  - \overline u(x))$ in the $\Sigma^+$, with
(dashed line) and without (dot-dashed line) Pauli blocking. 
All the curves were calculated in the MCM.

\item[{\bf Fig. 7}]   $x (d(x)  - \overline d(x))$ in the $\Sigma^+$ with the 
MCM. The octet and decuplet contributions are the dashed and the dotted lines,
respectively.

\item[{\bf Fig. 8}] a) $ x (s(x) - \overline s(x))$ in the proton (solid line) and 
$x (d(x)  - \overline d(x))$ in the $\Sigma^+$ (dashed line). Both curves
were calculated in the MCM.

\end{itemize}

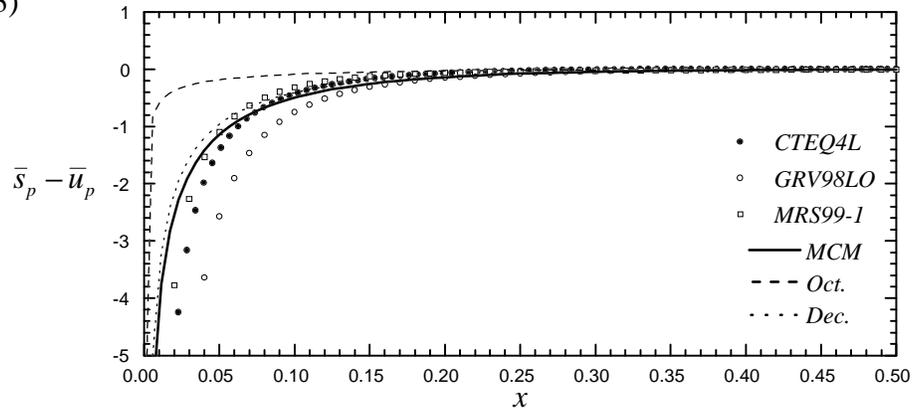
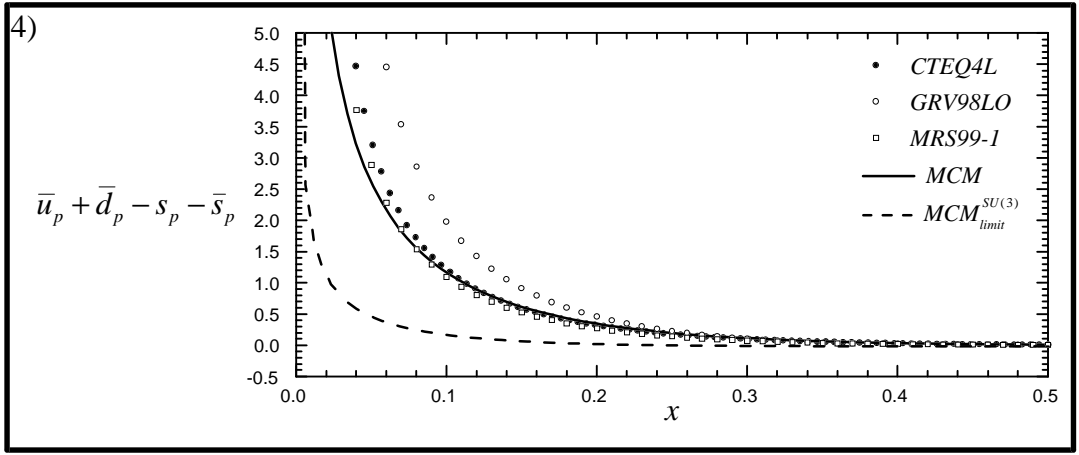
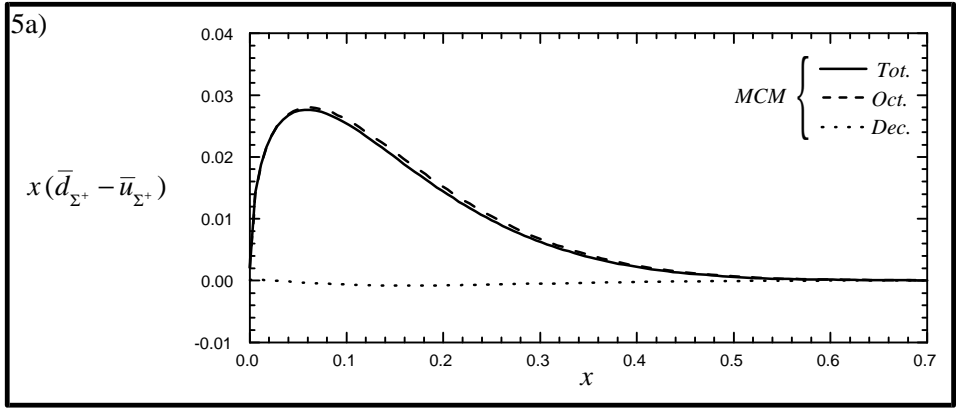
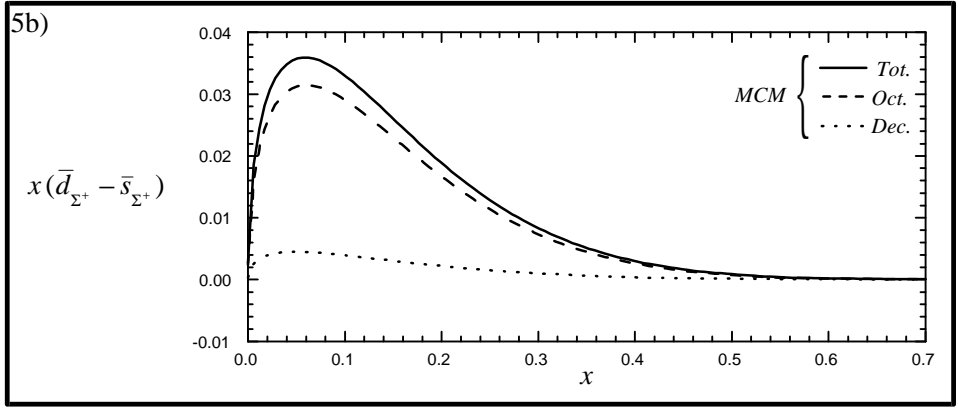
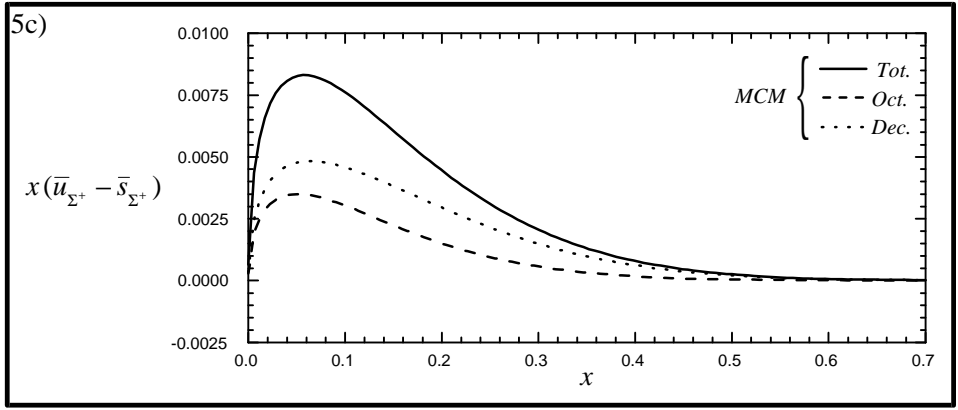
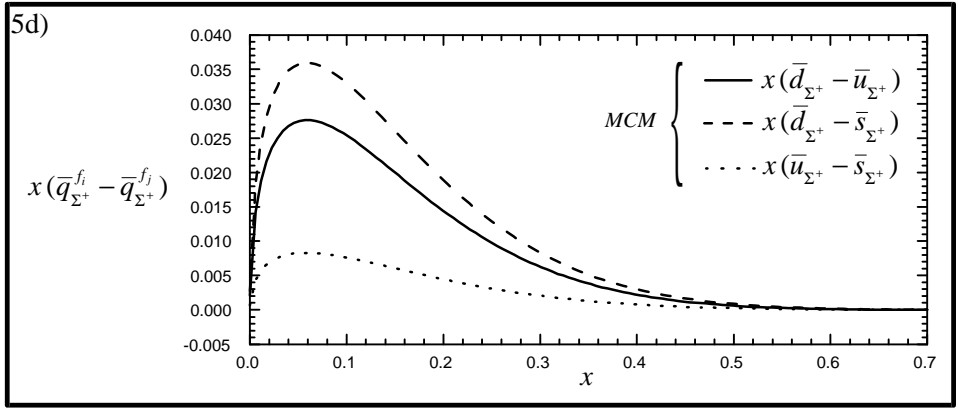
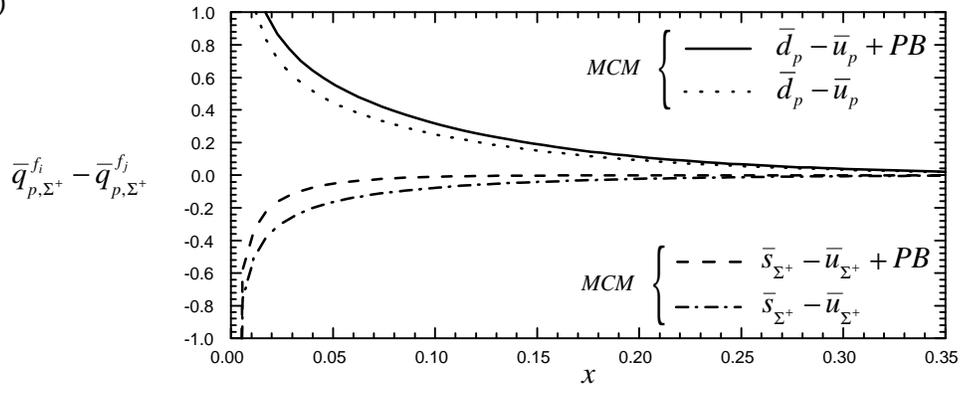
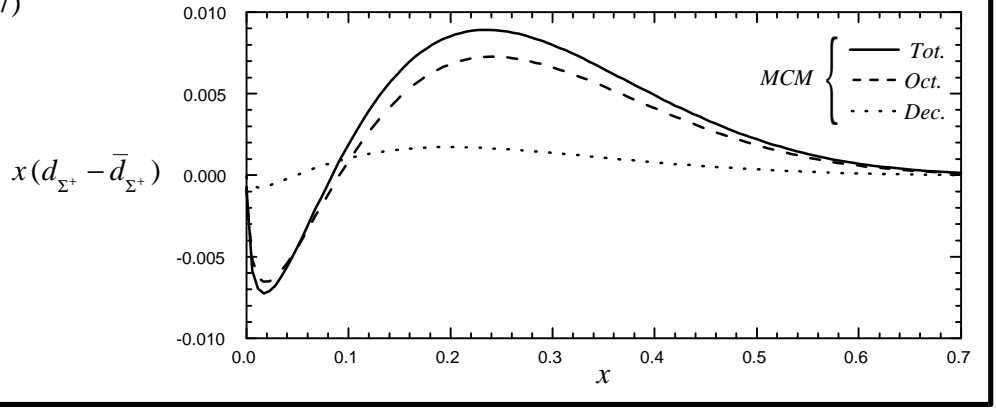
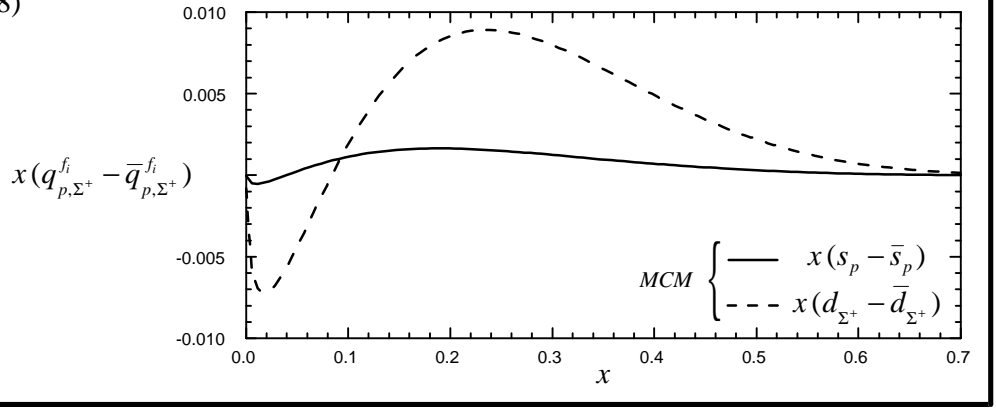
\end{document}